\documentclass[12pt,a4paper]{article}

\usepackage[usenames,dvipsnames]{xcolor}
\usepackage{lineno}
\usepackage{xspace}
\usepackage{scrextend}
\usepackage{dcolumn,booktabs}
\usepackage{xstring}
\usepackage{comment}
\usepackage{graphicx}
\usepackage{amsmath}
\usepackage[utf8]{inputenc}
\usepackage{textcomp}
\usepackage[normalem]{ulem}
\usepackage{multirow}
\usepackage{tabularx}

\usepackage{fp}
\usepackage{pgf}
\usepackage{siunitx}

\usepackage{hyperref}
\usepackage{url}
\usepackage[export]{adjustbox}

\newcommand{\tvf}{tvf-EMD}
\newcommand{\pytvf}{\texttt{pytvfemd}\xspace}
\newcommand{\tool}{\texttt{gwadaptive\_scattering}\xspace}

\newcommand{\validdate}{December 12, 2019 20:56:40}
\newcommand{\validgps}{1260219418}
\newcommand{\validduration}{40~\si{\second}}
\newcommand{\validcorrelation}{0.77}

\newcommand{\sqrthz}{\si{\per\sqrt{\text{\hertz}}}\xspace}

\newcommand{\TrigALLO}{$3706$\xspace}
\newcommand{\TrigBLLO}{$24566$\xspace}
\newcommand{\TrigCLLO}{$8623$\xspace}

\usepackage{chngcntr}
\counterwithout{figure}{section}

\begin{document}

\pagenumbering{arabic}
\title{{ \bf \tool: an automated pipeline for scattered light noise characterization}}

\author{Stefano Bianchi$^a$, Alessandro Longo$^{b}$, Guillermo Valdes$^{c,d}$ \\Gabriela Gonz\'alez$^c$, Wolfango Plastino$^{e,b}$ 
}

\date{\small \it{$^a$Rome, Italy}\\
\it{$^b$INFN, Sezione di Roma Tre, Via della Vasca Navale 84, 00146 Rome, Italy.}\\
\it{$^{c}$ Louisiana State University, Baton Rouge, Louisiana 70803, USA}\\
\it{$^{d}$ Texas A\&M University, College Station, Texas 77843, USA}\\
\it{$^e$Department of Mathematics and Physics, Roma Tre University, Via della Vasca Navale, 00146, Rome, Italy}
}
\maketitle

\begin{abstract}
Scattered light noise affects the sensitivity of gravitational waves detectors. The characterization of such noise is needed to mitigate it. The time-varying filter empirical mode decomposition algorithm is suitable for identifying signals with time-dependent frequency such as scattered light noise (or scattering). We present a fully automated pipeline based on the \pytvf library, a python implementation of the tvf-EMD algorithm, to identify objects inducing scattering in the gravitational-wave channel with their motion. The pipeline application to LIGO Livingston O3 data shows that most scattering noise is due to the penultimate mass at the end of the X-arm of the detector (EXPUM) and with a motion in the micro-seismic frequency range.
\end{abstract}

\section{Introduction}
\label{sec:Intro}
Advanced LIGO \cite{TheLIGOScientific:2014jea} and Advanced Virgo \cite{Losurdo:2017ais} are gravitational-wave detectors, instruments designed to measure tiny spacetime distortions called gravitational waves.
These extreme-sensitive instruments are capable of detecting strain amplitudes smaller than \num{e-23} \sqrthz and have observed gravitational waves from distinct systems such as binary black holes, binary neutron stars, and neutron star - black hole \cite{LIGOScientific:2018mvr,Abbott:2020niy,LIGOScientific:2021qlt}. However, such sensitivity can be overwhelmed by intrinsic noises or local environment noises coupling into the instrument \cite{TheLIGOScientific:2016zmo,Nuttall:2018xhi,Davis:2021ecd}. One common noise affecting the sensitivity of gravitational-wave detectors is scattered-light noise \cite{Soni:2021cjy}.  It originates when a fraction of laser light is undesirably scattered, reflected from moving surfaces, and recombined in any of the detectors' optical cavities \cite{Accadia:2010zzb}.  Scattered-light noise (or \textit{scattering}) manifests as arches in the detector's output with a frequency given by: 

\begin{equation}
f_{\text{arch}}(t)=\frac{2N}{\lambda}|v_{\text{surface}}(t)|
\label{eq:ffringe}
\end{equation}

Where $\lambda$ is the laser wavelength, $v_{\text{surface}}$ is the velocity at which the scattering surface is moving, and $N$ is the number of times the perturbed surface reflects the light before it recombines with the laser beam again. Studies had been made to locate and mitigate scattering in LIGO \cite{Valdes:2017xce,Soni:2020rbu}. These methods have employed the Hilbert-Huang transform \cite{1998RSPSA.454..903H}, a technique for analyzing nonlinear and non-stationary data, consisting of utilizing empirical mode decomposition (EMD) and the Hilbert transform. In Virgo, the time-varying filter EMD (tvf-EMD) \cite{Li_2017} adaptive algorithm was used to improve the scattering's origin localization \cite{Longo:2020onu} and to denoise environmental signals \cite{Longo:2020wpg}.
Posteriorly, the scattering characterization, e.g., identifying the reflective surface motion frequency, is necessary to understand how the noise originated. For scattered light characterization, we used \pytvf, a Python library that implements tvf-EMD \cite{Bianchi:pytvfemd}. We show results from the application of the tool on data recorded by the LIGO Livingston detector (LLO) during the third observing run (so called O3). We discuss which areas of the detector and frequencies had a high correlation with the scattering, based on the adopted methodology. Finally, we present \tool, a new fully automated pipeline based on tvf-EMD to locate and characterize the origin of scattered light glitches as obtained from GravitySpy, a machine learning image recognition tool that classifies transient noise instances (also called triggers) in the strain data \cite{Zevin:2016qwy}

\section{Methodology}
\label{sec:Methods}
\subsection{Time-varying filter EMD and \pytvf} 
\label{subsec:tvfEMD}
EMD is a method to empirically break down any nonlinear, non-stationary data set into a finite number of intrinsic mode functions (IMF) through a sifting process.
Ideally, each IMF will include a narrow band of oscillations, in order for Hilbert spectral analysis to give meaningful results. Still, EMD fails to distinguish components whose frequencies are too similar \cite{4359551}. Also, noisy data could lead to mode mixed IMFs, meaning that the IMF extracted by EMD might contain oscillations of widely different scales or that multiple IMFs contain the same oscillatory mode \cite{doi:10.1146/annurev.fluid.31.1.417}. A mode mixed IMF typically has no physical meaning. The time-varying filter EMD mitigates mode mixing and improves the IMFs frequency resolution, replacing the known IMF concept with the local narrow-band oscillatory modes. The tvf-EMD employs a B-spline \cite{799930} as the time-varying filter to extract these local narrow-band modes. Furthermore, the algorithm's sifting process stops when the ratio  between the Loughlin instantaneous bandwidth $B_{\text{Loughlin}}(t)$ and the weighted average of the instantaneous frequencies $\varphi_{\text{average}}(t)$ is lower than a selected threshold $\xi$. The definition of $B_{\text{Loughlin}}(t)$, $\varphi_{\text{average}}(t)$, and a full description of the algorithm can be found in \cite{Li_2017}. Regarding tvf-EMD, the relevant parameters to consider are the bandwitdth threshold ratio $\xi$ and the B-spline order $n$. While $\xi$ determines the threshold on the bandwidth of the extracted modes to stop the sifting process, $n$ is related to the time varying filter frequency roll off. Following \cite{Li_2017}, these parameters were set to $\xi=0.1$ and $n=26$. \pytvf \cite{Bianchi:pytvfemd} is a Python package that implements the \tvf ~algorithm based on an existing MATLAB code \cite{tvfemd_matlab}. In this paper it was used for scattering characterization, but it is also available for the analysis of other non linear and non stationary data. The main function \textit{tvfemd} takes the signal as input and returns its intrinsic mode functions. Other inputs are the algorithm's main parameters (i.e., $\xi$ and $n$) and the number of IMFs to be extracted. 

\subsection{Scattering characterization pipeline}
\label{subsec:char_pipeline}
The pipeline follows the same procedure described in \cite{Valdes:2017xce}, except that instead of EMD, we use tvf-EMD for the sifting process, as implemented in \cite{Longo:2020onu}.  The pipeline completes the following steps to locate the origin of scattering noise and characterize it:
\begin{enumerate}
	\item Identify times when the data is affected by scattering, using the times and peak frequencies of the scattering arches provided by GravitySpy. \label{st:gspy}
	\item Lowpass filter the data using the peak frequency as a cutoff. \label{st:lowp}
	\item Get the scattering's instantaneous amplitude \textit{IA}, using tvf-EMD and the Hilbert transform. \label{st:hht}
	\item Generate the scattering predictors for a given list of auxiliary sensors, using Equation 1. \label{st:pred}
	\item Estimate the Pearson correlation coefficient $\rho$ between the instantaneous amplitude and the predictors. We consider only the first IMF for each channel, as it is the one relevant after the lowpass filter. We call \textit{culprit} the predictor with the highest correlation. \label{st:corr}
	\item Estimate the associated frequency to the scattering using the culprit's motion frequency. We calculated this frequency by looking for the highest peak in the amplitude spectrum of the culprit's motion data.  \label{st:freq}
\end{enumerate}

Steps \ref{st:gspy} and \ref{st:freq} are an addition to this new version of the pipeline. For Step \ref{st:pred}, we use $N=1$ to calculate the scattering predictors because the correlation in Step \ref{st:corr} ignores the signals' amplitude. We use data from the third observing run (O3) for our analyses.  O3 began in April 2019 and lasted nearly one year, with a month-long commissioning break in October 2019.  The periods before and after the break are known as O3a and O3b, respectively.  In January 2020, scattering noise decreased after commissioners reduced the relative motion between the test mass and the reaction mass --see section 6 in \cite{Soni:2020rbu}. This noise reduction using the suspensions control system is known as \textit{RC tracking}. We identify these periods before and after the RC tracking implementation as O3b1 and O3b2. 

We select the triggers classified as \textit{scattered\_light} by GravitySpy, with a confidence greater than 0.9 and SNR greater than 10. There are \TrigALLO, \TrigBLLO, and \TrigCLLO triggers in LIGO Livingston data with these characteristics in O3a, O3b1, and O3b2, respectively, after removing triggers with duplicated GPS time. Table \ref{tab:dataset} summarizes the scattering triggers of the study. 

\begin{table}[ht]
\caption{Data employed in this study. Scattering triggers with confidence greater than 0.9 and SNR greater than 10, during O3 in LIGO Livingston.}
\begin{center}
\resizebox{\textwidth}{!}{
\begin{tabular}{@{}lllll@{}}
\toprule
Epoch & Starting date & Ending date  & Description                       & \begin{tabular}[c]{@{}l@{}}Scattering\\ triggers\end{tabular} \\ \midrule
O3a   & Apr 01, 2019  & Sep 30, 2019 & O3 first part                     & \TrigALLO                                                          \\
Comm  & Oct 01, 2019  & Oct 31, 2019 & O3 commissioning break            & NA                                                            \\
O3b1  & Nov 01, 2019  & Jan 06, 2020 & O3 second part before RC tracking & \TrigBLLO                                                         \\
O3b2  & Jan 07, 2020  & Mar 27, 2020 & O3 second part after RC tracking  & \TrigCLLO                                                          \\ \bottomrule
\end{tabular}
}
\label{tab:dataset}
\end{center}
\end{table}

LIGO samples the gravitational-wave channel data (so-called DARM-Differential Arm Motion), with a frequency of 16384~\si{\hertz}, and auxiliary data such as the position of the detector’s mirrors at lower rates. We use 60~\si{\second} of DARM and auxiliary data around the peak time of each scattering trigger and downsampled at 256~\si{\hertz}. Besides the gravitational-wave channel, the LIGO detectors record over 200000 auxiliary sensors (or channels) that monitor instrument behavior and environmental conditions. We use a short list of channels monitoring the optics' suspension system top stage motion for our study, including the channel monitoring the end X-arm penultimate stage motion. We use DARM and auxiliary channels to calculate the scattering instantaneous amplitude and the scattering predictors, respectively. In Appendix \ref{ap:list}, we list the name of these channels, including a short description. In Appendix \ref{ap:layout}, we include a LIGO configuration schematic to visualize the location of its optics better. A list of abbreviations and acronyms can be consulted in \cite{Acronyms_2020}.

\section{Results}
\label{sec:Results}
\subsection{Validation}
\label{subsec:validation}

\begin{figure}[!ht]
	\begin{center}
		\includegraphics[width=0.495\columnwidth, height=4.9cm]{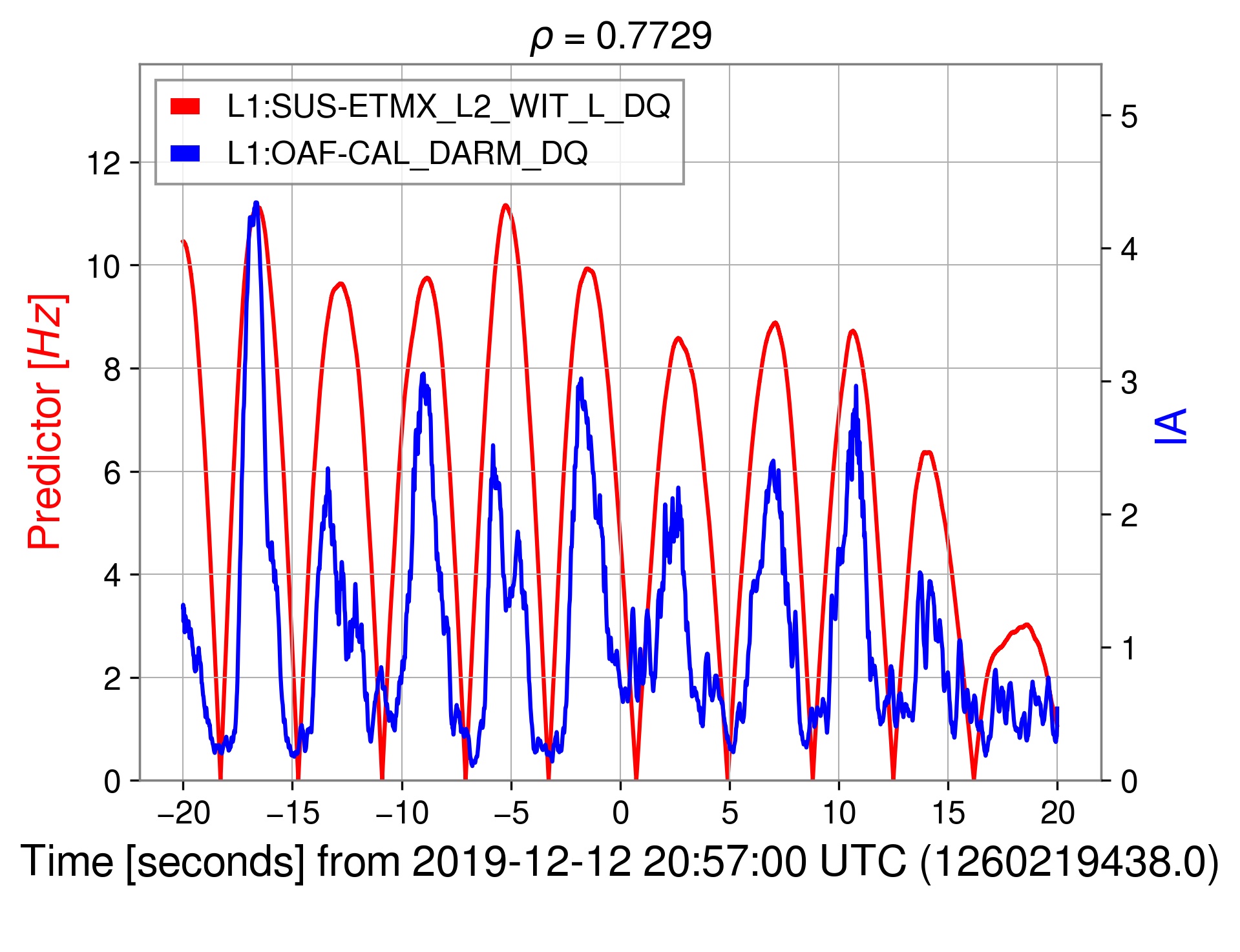}
		\includegraphics[width=0.495\columnwidth]{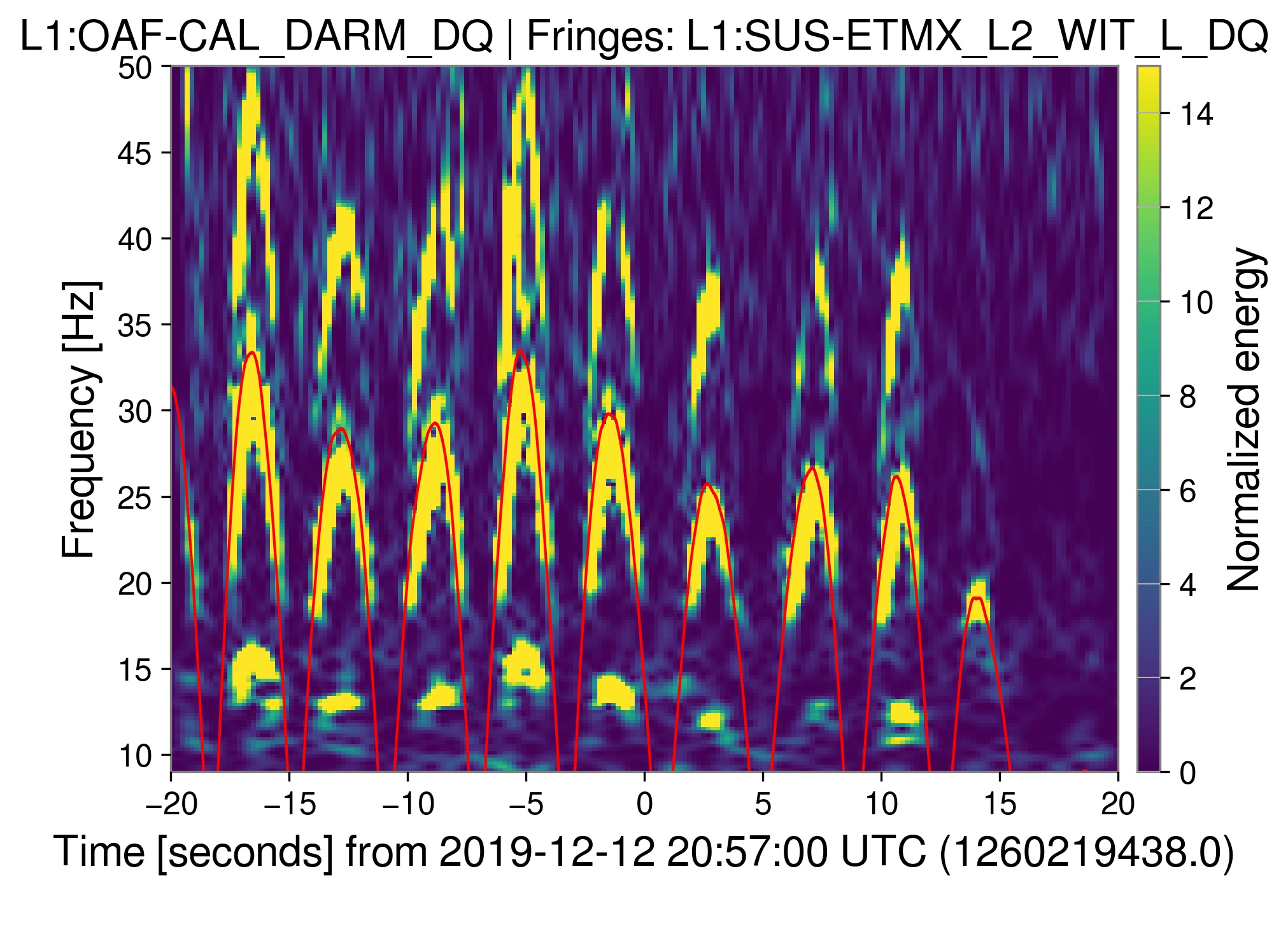}
		\caption{Instantaneous amplitude (left plot) and spectrogram (right plot) of DARM data dominated by scattering noise and the scattering culprit (red line). In this validation example, the culprit is associated with the end X-arm penultimate mass (EXPUM). We used $N=3$ to generate the EXPUM scattering predictor for better visualization of the overlapping.}
	\end{center}
	\label{fig:validation}
\end{figure}

We tested the tool on data dominated by scattering noise previously identified to be correlated with the motion of penultimate mass at the end of the X arm of the detector (EXPUM) \cite{alog:schofield_scattering}, with starting time \validdate~(GPS \validgps). 
We used the list of channels in Table \ref{tab:scat_list}, and the tool pointed EXPUM as the culprit with $\rho =$~\validcorrelation. 
Other end X-arm predictors had correlations $\rho < 0.73$. In contrast, predictors unrelated to the end X-arm had correlations $\rho < 0.39$. 
Figure \ref{fig:validation} shows the culprit, i.e. the EXPUM predictor, overlapping with the DARM instantaneous amplitude (left graph) and spectrogram (right graph), validating the performance of our tool. We generated the EXPUM predictor, \validduration~around GPS 1260219438, with $N = 3$ for better overlapping visualization.

\subsection{Scattering associated to X-arm end mass}
\label{subsec:tracking}
The implementation of RC tracking between O3b1 and O3b2 lead to a reduction of scattered light glitches due to the reaction mass \cite{Soni:2020rbu}. 
With our method, we also found a reduction in the percentage of scattered light glitches for which the obtained culprit is EXPUM, with $\rho > 0.5$, between O3b1 and O3b2.

We grouped the associated frequency to the scattering in the bands defined in Table \ref{tab:bands}. We also found that the percentage of EXPUM scattering associated with the frequency band B decreased in O3b2. Our results support those in \cite{Soni:2020rbu}, where they found that a major part of the scattering was associated with the 0.1~\si{\hertz}--0.3~\si{\hertz} band (so-called micro-seismic band) and was reduced after the RC tracking implementation.

\begin{table}[ht]
\caption{Correspondence between frequency bands values and labels.}
\begin{center}
\begin{tabular}{cl|cl}
\toprule
Label & Frequency band [\si{\hertz}] & Label & Frequency band [\si{\hertz}] \\
\midrule
A & 0.03--0.10 & D & 1.00--3.00 \\
B & 0.10--0.30 & E & 3.00--10.0 \\
C & 0.30--1.00 & ELSE & $<$ 0.03, $\geq$ 10.0 \\
\bottomrule
\end{tabular}
\label{tab:bands}
\end{center}
\end{table}

\begin{figure}[ht!]
	\begin{center}
		\includegraphics[width=0.495\columnwidth]{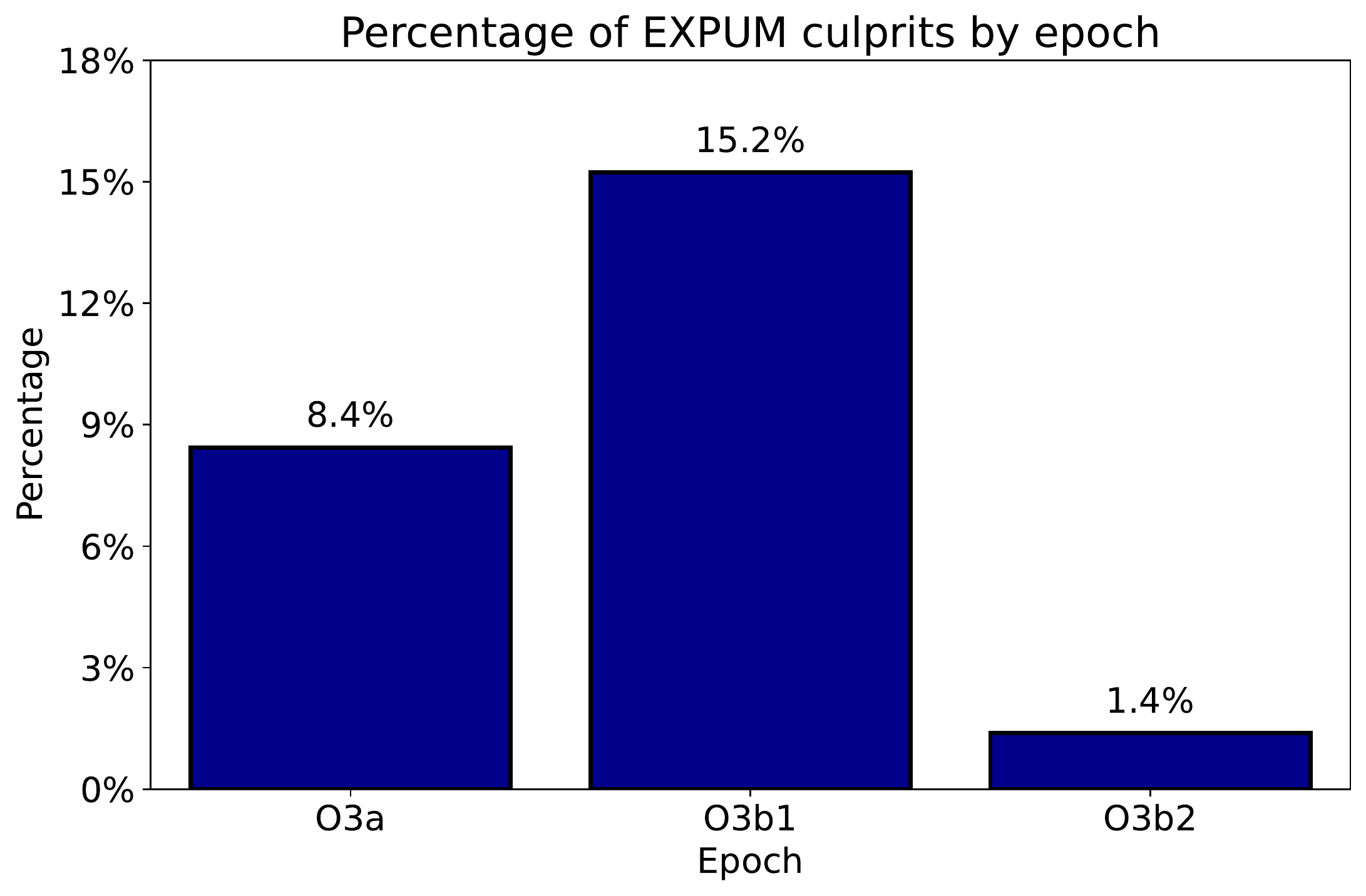}
		\includegraphics[width=0.495\columnwidth]{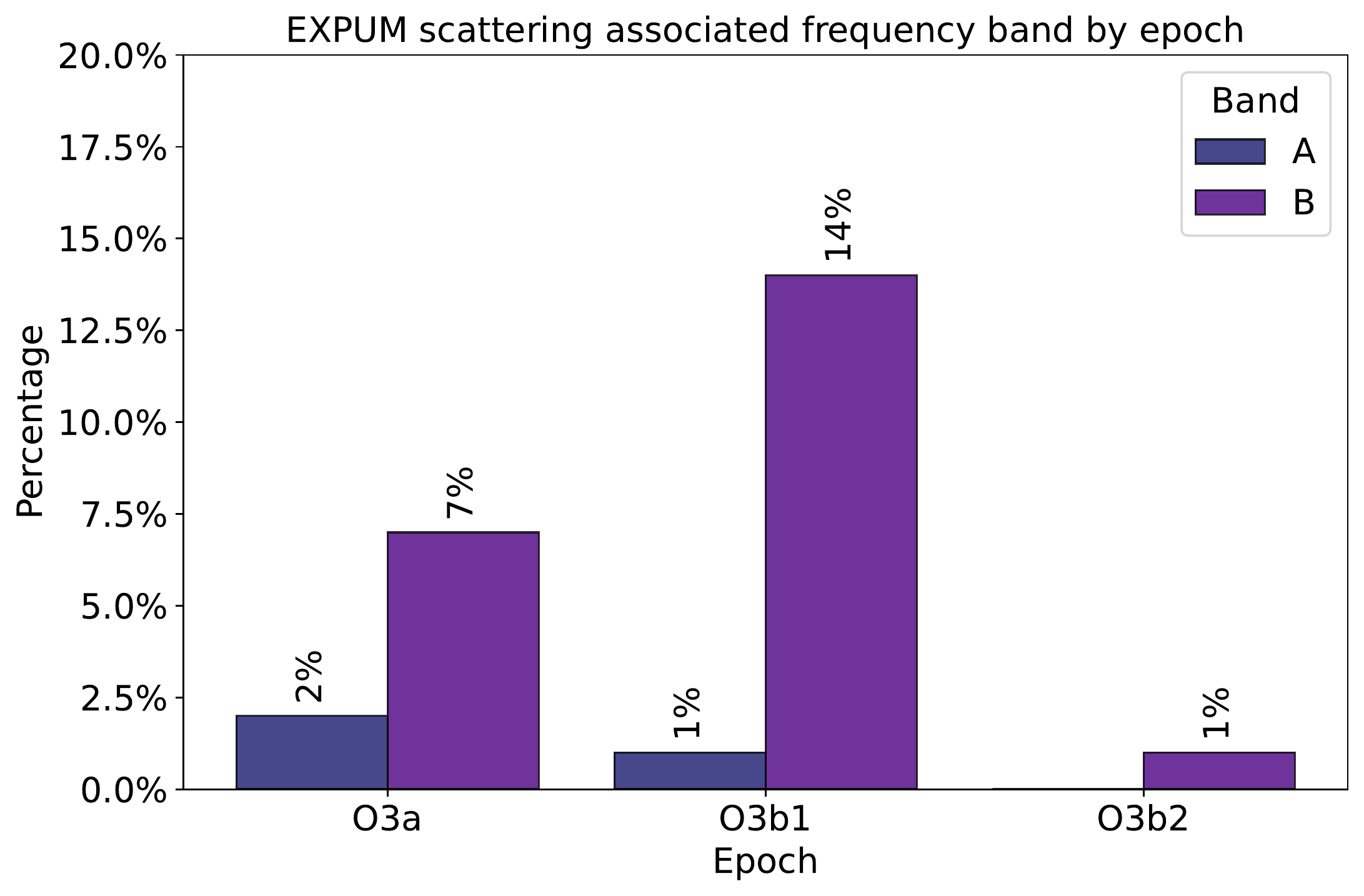}
		\caption{Left plot - Percentage of scattering in LIGO Livingston associated with EXPUM during O3.  Right plot - Percentage of scattering in LIGO Livingston associated with EXPUM and the frequency bands A (0.03~\si{\hertz}--0.1~\si{\hertz}) and B (0.1~\si{\hertz}--0.3~\si{\hertz}) during O3. Our tool quantified the reduction of this scattering after commissioning improvements. }  
	\end{center}
	\label{fig:scattering_ERMX}
\end{figure}

Figure \ref{fig:scattering_ERMX} shows that the percentage of EXPUM scattering decreased from 15.2\% in O3b1 to 1.4\% in O3b2 (left plot) and the EXPUM scattering associated with the micro-seismic band decreased from 14\% in O3b1 to 1\% in O3b2 (right plot).



\subsection{O3 scattering}
\label{subsec:bands}
We extended the study to investigate the scattering origin and associated frequency based on location within the detector. 
Consult Figure \ref{fig:cartoon} for better comprehension of the distinct vacuum chamber locations.

\begin{figure}[ht]
	\begin{center}
		\includegraphics[width=0.495\columnwidth]{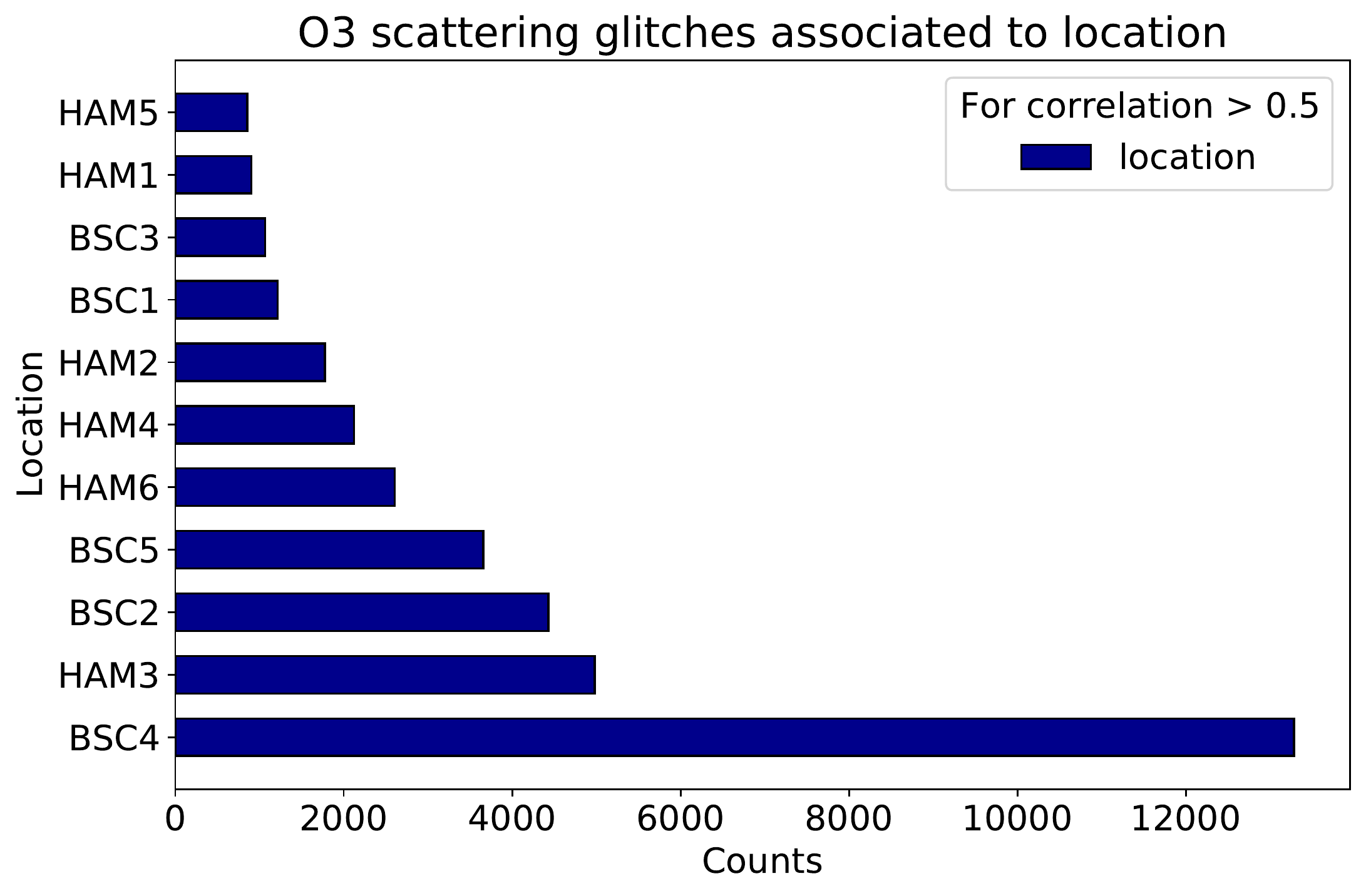}
		\includegraphics[width=0.495\columnwidth]{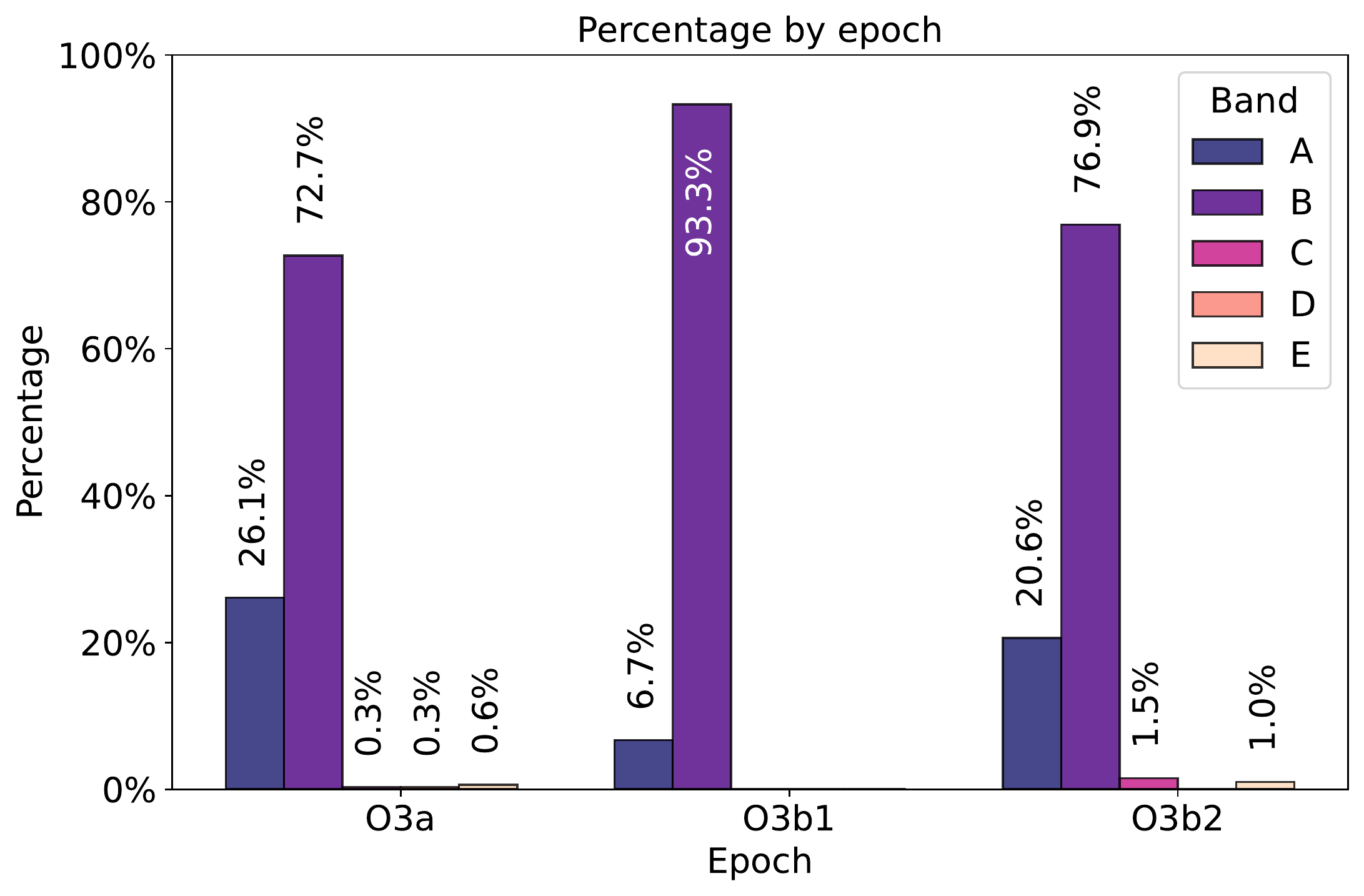}
		\caption{Left plot - Ranking of O3 scattering by location in the detector. Right plot - Percentage of scattering associated with the distinct frequency bands in Table \ref{tab:bands} during O3. We omit the columns with a value equals to 0.}
	\end{center}
	\label{fig:O3_scattering}
\end{figure}

We found that most of the O3 scattering in LIGO Livingston was linked to BSC4, the vacuum chamber enclosing the X-arm end test mass, and therefore EXPUM.  When grouped by the scattering associated frequency, we observed a reduction in the band B scattering,  from 93.3\% in O3b1 to 76.9\% in O3b2. We saw a reduction of scattering in band A in O3b2 with respect to O3a, but an increase with respect to O3b1.  Figure \ref{fig:O3_scattering} shows the ranking of O3 scattering by location in the detector (left plot) and the percentage of scattering associated with the frequency bands by epoch (right plot). 



\section{Daily-basis automated pipeline} 
\label{sec:Pipeline}

\begin{figure}[ht!]
	\centering
	\includegraphics[width=\columnwidth]{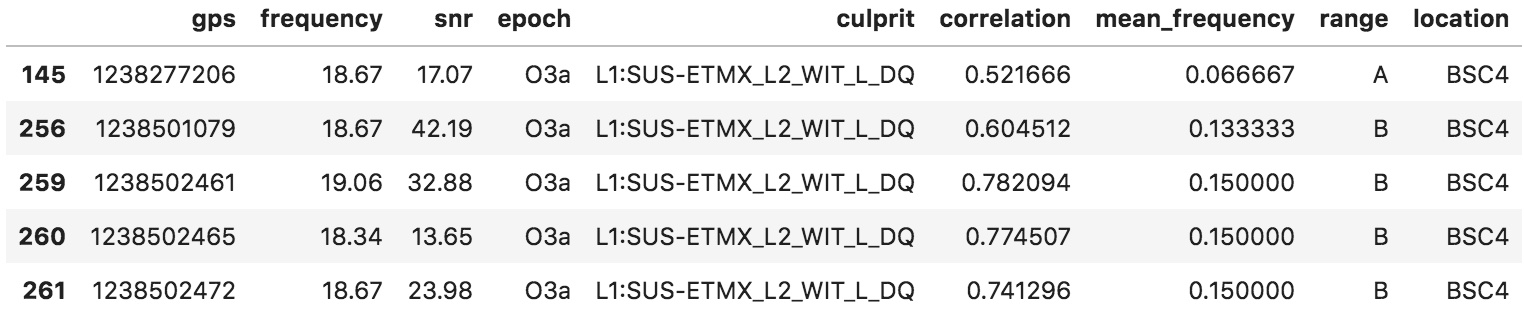}
	\caption{Screenshot of the results data frame generated by our \tool pipeline.}
	\label{fig:screenshot}
\end{figure}

We present \tool, a Python package that builds the automated pipeline for the scattering localization and characterization on the LIGO computers.
The pipeline generates three main files for HTCondor \cite{Bockelman_2015}, a high-throughput computing software for distributed parallelization of tasks to accelerate the process for a given target channel (i.e., the scattering witness channel), and GPS times list. The first file completes the steps describes in Section \ref{subsec:char_pipeline}. We called these the analysis jobs. The second file merges the analysis jobs results in a single data frame. We denominate these the summary jobs. The third file organizes the hierarchy, running the analysis jobs first and then the summary jobs.  A screenshot of the summary jobs data frame is shown in Figure \ref{fig:screenshot} and includes the following columns:


\begin{itemize}
	\item GPS: GPS time (from GravitySpy).
	\item Frequency: Glitch peak frequency (from GravitySpy).
	\item SNR: Glitch signal-to-noise ratio (from GravitySpy).
	\item Epoch: O3a, O3b1, O3b2.
	\item Culprit: Name of channel with highest correlation. For possible values see Table \ref{tab:scat_list}.
	\item Correlation: Correlation coefficient between scattering and culprit.
	\item Mean frequency: Culprit's mean frequency calculated as the highest peak in its spectrum.
	\item Range: Ground motion frequency bands used in LIGO. Possible values are A, B, C, D, E, and ELSE (see Table \ref{tab:bands}).
	\item Location: Chamber containing the optic (see Table \ref{tab:scat_list}).
\end{itemize}

Additionally, the pipeline creates a directory containing the results and the plots and presents them in an HTML page format.

%


\section{Conclusions}
\label{sec:Conclusions}
We have presented an automated pipeline based on the tvf-EMD algorithm to identify and characterize scattered light noise in data recorded by the LIGO Livingston gravitational-wave detector in its third observing run, O3. 


We found the correlation between the LIGO Livingston scattering noise, the motion of the penultimate mass at the end of the detector's X-arm, and the micro-seismic frequency band (0.1~\si{\hertz}--0.3~\si{\hertz}) was high; a similar result previously discovered by other means. We used our tool to quantify the percentage of scattering by location and frequency. We found that most of the scattering was linked to the chamber enclosing the X-arm end mass, and the micro-seismic frequency scattering decreased after the commissioning improvements known as RC tracking.

In upcoming observing runs, this tool can be used in other detectors such as LIGO Hanford and Virgo. Its results could be presented in summary pages for rapid consultation aimed to help commissioners. The long-term analysis of the pipeline results will lead to a better characterization of the detectors.


The \tool pipeline proved to be useful in identifying and characterizing time- and frequency-varying features in data of large-scale gravitational-wave detectors. The adaptive nature of the tvf-emd algorithm to nonlinear and non-stationary data suggests that our tool could be employed to identify and characterize another type of noise in gravitational-wave astronomy or other fields.

\section*{Acknowledgements}

LIGO was constructed by the California Institute of Technology and Massachusetts Institute of Technology with funding from the National Science Foundation, and it operates under Cooperative Agreement No. PHY-1764464. 
Advanced LIGO was built under Grant No. PHY-0823459. 
We use the LIGO Livingston detector data for this paper and the LIGO computing resources to perform this work.
The \pytvf algorithm, fundamental for this work, was developed in the Virgo Collaboration framework.
We acknowledge the discussions with members of the Detector Characterization Group of the LIGO Scientific and Virgo Collaboration. 
We thank LIGO Scientific Collaboration members B. O'Reilly and C. Berry for their helpful feedback.
This research was aided by the knowledge and skills acquired through participation in the Open Science Grid (OSG) User School,  supported  by the National Science Foundation under Grant No.  PHY-1148698.
G. Valdes thank Consejo Nacional de Ciencia y Tecnolog\'ia -  CONACyT of M\'exico for the support provided through the assistantship \textit{Estancias Posdoctorales en el Extranjero}.


\bibliographystyle{ieeetr}

\begin{thebibliography}{10}

\bibitem{TheLIGOScientific:2014jea}
J.~Aasi {\em et~al.}, ``{Advanced LIGO},'' {\em Class. Quant. Grav.}, vol.~32,
  p.~074001, 2015.

\bibitem{Losurdo:2017ais}
F.~Acernese, T.~Adams, K.~Agatsuma, L.~Aiello, A.~Allocca, A.~Amato, S.~Antier,
  N.~Arnaud, S.~Ascenzi, P.~Astone, {\em et~al.}, ``{Status of the Advanced
  Virgo gravitational wave detector},'' {\em International Journal of Modern
  Physics A}, vol.~32, no.~28n29, p.~1744003, 2017.

\bibitem{LIGOScientific:2018mvr}
B.~P. Abbott {\em et~al.}, ``{GWTC-1: A Gravitational-Wave Transient Catalog of
  Compact Binary Mergers Observed by LIGO and Virgo during the First and Second
  Observing Runs},'' {\em Phys. Rev. X}, vol.~9, no.~3, p.~031040, 2019.

\bibitem{Abbott:2020niy}
R.~Abbott {\em et~al.}, ``{GWTC-2: Compact binary coalescences observed by LIGO
  and Virgo during the first half of the third observing run},'' {\em Phys.
  Rev. X}, vol.~11, no.~2, p.~021053, 2021.

\bibitem{LIGOScientific:2021qlt}
R.~Abbott {\em et~al.}, ``{Observation of Gravitational Waves from Two Neutron
  Star\textendash{}Black Hole Coalescences},'' {\em Astrophys. J. Lett.},
  vol.~915, no.~1, p.~L5, 2021.

\bibitem{TheLIGOScientific:2016zmo}
B.~P. Abbott {\em et~al.}, ``{Characterization of transient noise in Advanced
  LIGO relevant to gravitational wave signal GW150914},'' {\em Class. Quant.
  Grav.}, vol.~33, no.~13, p.~134001, 2016.

\bibitem{Nuttall:2018xhi}
L.~Nuttall, ``{Characterizing transient noise in the LIGO detectors},'' {\em
  Philosophical Transactions of the Royal Society A: Mathematical, Physical and
  Engineering Sciences}, vol.~376, no.~2120, p.~20170286, 2018.

\bibitem{Davis:2021ecd}
D.~Davis, J.~S. Areeda, B.~K. Berger, R.~Bruntz, A.~Effler, R.~C. Essick, R.~P.
  Fisher, P.~Godwin, E.~Goetz, A.~F. Helmling-Cornell, {\em et~al.}, ``{LIGO
  detector characterization in the second and third observing runs},'' {\em
  Class. Quant. Grav.}, vol.~38, no.~13, p.~135014, 2021.

\bibitem{Soni:2021cjy}
S.~Soni {\em et~al.}, ``{Discovering features in gravitational-wave data
  through detector characterization, citizen science and machine learning},'' 3
  2021.

\bibitem{Accadia:2010zzb}
T.~Accadia {\em et~al.}, ``{Noise from scattered light in Virgo's second
  science run data},'' {\em Class. Quant. Grav.}, vol.~27, p.~194011, 2010.

\bibitem{Valdes:2017xce}
G.~Valdes, B.~O'Reilly, and M.~Diaz, ``{A Hilbert\textendash{}Huang transform
  method for scattering identification in LIGO},'' {\em Class. Quant. Grav.},
  vol.~34, no.~23, p.~235009, 2017.

\bibitem{Soni:2020rbu}
S.~Soni {\em et~al.}, ``{Reducing scattered light in LIGO's third observing
  run},'' {\em Class. Quant. Grav.}, vol.~38, no.~2, p.~025016, 2020.

\bibitem{1998RSPSA.454..903H}
N.~E. {Huang}, Z.~{Shen}, S.~R. {Long}, M.~C. {Wu}, H.~H. {Shih}, Q.~{Zheng},
  N.~C. {Yen}, C.~C. {Tung}, and H.~H. {Liu}, ``{The empirical mode
  decomposition and the Hilbert spectrum for nonlinear and non-stationary time
  series analysis},'' {\em Proceedings of the Royal Society of London Series
  A}, vol.~454, pp.~903--998, March 1998.

\bibitem{Li_2017}
H.~Li, Z.~Li, and W.~Mo, ``A time varying filter approach for empirical mode
  decomposition,'' {\em Signal Processing}, vol.~138, 03 2017.

\bibitem{Longo:2020onu}
A.~Longo, S.~Bianchi, W.~Plastino, N.~Arnaud, A.~Chiummo, I.~Fiori,
  B.~Swinkels, and M.~Was, ``{Scattered light noise characterisation at the
  Virgo interferometer with tvf-EMD adaptive algorithm},'' {\em Class. Quant.
  Grav.}, vol.~37, no.~14, p.~145011, 2020.

\bibitem{Longo:2020wpg}
A.~Longo, S.~Bianchi, W.~Plastino, I.~Fiori, D.~Fiorucci, J.~Harms,
  F.~Paoletti, M.~Barsuglia, and M.~Falxa, ``{Adaptive Denoising of Acoustic
  Noise Injections Performed at the Virgo Interferometer},'' {\em Pure Appl.
  Geophys.}, vol.~177, no.~7, pp.~3395--3406, 2020.

\bibitem{Bianchi:pytvfemd}
S.~Bianchi, ``stfbnc/pytvfemd v0.1,'' {\em Zenodo}, February 2021.
\newblock https://doi.org/10.5281/zenodo.4568706.

\bibitem{Zevin:2016qwy}
M.~Zevin {\em et~al.}, ``{Gravity Spy: Integrating Advanced LIGO Detector
  Characterization, Machine Learning, and Citizen Science},'' {\em Class.
  Quant. Grav.}, vol.~34, no.~6, p.~064003, 2017.

\bibitem{4359551}
G.~{Rilling} and P.~{Flandrin}, ``One or two frequencies? the empirical mode
  decomposition answers,'' {\em IEEE Transactions on Signal Processing},
  vol.~56, no.~1, pp.~85--95, 2008.

\bibitem{doi:10.1146/annurev.fluid.31.1.417}
N.~E. Huang, Z.~Shen, and S.~R. Long, ``A new view of nonlinear water waves:
  The {H}ilbert spectrum,'' {\em Annual Review of Fluid Mechanics}, vol.~31,
  no.~1, pp.~417--457, 1999.

\bibitem{799930}
M.~{Unser}, ``Splines: a perfect fit for signal and image processing,'' {\em
  IEEE Signal Processing Magazine}, vol.~16, no.~6, pp.~22--38, 1999.

\bibitem{tvfemd_matlab}
S.~Li, ``Time varying filter based empirical mode decomposition (tvf-emd),''
  {\em MATLAB Central File Exchange}, 2021.

\bibitem{Acronyms_2020}
D.~Shoemaker {\em et~al.}, ``{LIGO} - {V}irgo - {KAGRA} ({LVK}) {A}bbreviations
  and {A}cronyms {L}ist.'' LIGO-Document-M080375, 2020.
\newblock https://dcc.ligo.org/LIGO-M080375/public.

\bibitem{alog:schofield_scattering}
R.~Schofield, ``{aLIGO LHO Logbook}.''
  \href{https://alog.ligo-wa.caltech.edu/aLOG/index.php?callRep=54298}{54298},
  2020.

\bibitem{Bockelman_2015}
B.~Bockelman, T.~Cartwright, J.~Frey, E.~M. Fajardo, B.~Lin, M.~Selmeci,
  T.~Tannenbaum, and M.~Zvada, ``Commissioning the {HTCondor}-{CE} for the open
  science grid,'' {\em Journal of Physics: Conference Series}, vol.~664,
  p.~062003, December 2015.

\bibitem{Kissel_2017}
J.~Kissel, ``a{LIGO} {S}eismic {I}solation and {S}uspensions cartoon.''
  LIGO-Document-G1200071, 2017.
\newblock https://dcc.ligo.org/LIGO-G1200071/public.

\bibitem{matichard2016overview}
F.~Matichard, D.~Martynov, B.~Shapiro, J.~Rollins, and D.~Coyne, ``An overview
  of the control layers in ligo 4km interferometers,'' in {\em Proceedings-ASPE
  2016 Spring Topical Meeting: Precision Mechatronic System Design and
  Control}, pp.~17--22, 2016.

\end{thebibliography}

\pagebreak
\appendix
\section{Channel names} 
\label{ap:list}

\begin{table}[ht]
\caption{\label{app:scat_list} List of channels of investigated objects, including 23 mirrors, 4 reaction masses (CPX, CPY, ERMX, ERMY), and 2 transmission monitors (TMSX, TMSY). Where L refers to the longitudinal directions.} 
\footnotesize\rm
\begin{center}
\resizebox{0.99\textwidth}{!}{
\begin{tabular}{@{}llrl}
\toprule
Short&	Channel&	Chamber&	Description\\
Name&		   &	\\
\midrule
DARM&		L1:OAF-CAL\_DARM\_DQ&	NA&	Gravitational-wave channel\\
\midrule
BS&		L1:SUS-BS\_M2\_WIT\_L\_DQ&		BSC2&	Beam Splitter\\


MC1&	L1:SUS-MC1\_M1\_DAMP\_L\_IN1\_DQ&	HAM2&	Mode Cleaner optic 1\\
MC2&	L1:SUS-MC2\_M1\_DAMP\_L\_IN1\_DQ&	HAM3&	Mode Cleaner optic 2\\
MC3&	L1:SUS-MC3\_M1\_DAMP\_L\_IN1\_DQ&	HAM2&	Mode Cleaner optic 3\\
PRM&	L1:SUS-PRM\_M1\_DAMP\_L\_IN1\_DQ&	HAM2&	Power Recycling Mirror\\
PR2&	L1:SUS-PR2\_M1\_DAMP\_L\_IN1\_DQ&	HAM3&	Power Recycling optic 2\\
PR3&	L1:SUS-PR3\_M1\_DAMP\_L\_IN1\_DQ&	HAM2&	Power Recycling optic 3\\
SRM&	L1:SUS-SRM\_M1\_DAMP\_L\_IN1\_DQ&	HAM5&	Signal Recycling Mirror\\
SR2&	L1:SUS-SR2\_M1\_DAMP\_L\_IN1\_DQ&	HAM4&	Signal Recycling optic 2\\
SR3&	L1:SUS-SR3\_M1\_DAMP\_L\_IN1\_DQ&	HAM5&	Signal Recycling optic 3\\
RM1&	L1:SUS-RM1\_M1\_DAMP\_L\_IN1\_DQ&	HAM1&	Recycling Mirror 1\\
RM2&	L1:SUS-RM2\_M1\_DAMP\_L\_IN1\_DQ&	HAM1&	Recycling Mirror 2\\
IM1&		L1:SUS-IM1\_M1\_DAMP\_L\_IN1\_DQ&	HAM2&	Input Mirror 1\\
IM2&		L1:SUS-IM2\_M1\_DAMP\_L\_IN1\_DQ&	HAM2&	Input Mirror 2\\
IM3&	L1:SUS-IM3\_M1\_DAMP\_L\_IN1\_DQ&	HAM2&	Input Mirror 3\\
IM4&	L1:SUS-IM4\_M1\_DAMP\_L\_IN1\_DQ&	HAM2&	Input Mirror 4\\
OM1&	L1:SUS-OM1\_M1\_DAMP\_L\_IN1\_DQ&	HAM6&	Output Mirror 1\\
OM2&	L1:SUS-OM2\_M1\_DAMP\_L\_IN1\_DQ&	HAM6&	Output Mirror 2\\
OM3&	L1:SUS-OM3\_M1\_DAMP\_L\_IN1\_DQ&	HAM6&	Output Mirror 3\\


TMSX&	L1:SUS-TMSX\_M1\_DAMP\_L\_IN1\_DQ&	BSC4&	Transmission Monitor\\&&& Suspension in X-arm\\
TMSY&	L1:SUS-TMSY\_M1\_DAMP\_L\_IN1\_DQ&	BSC5&	Transmission Monitor\\&&& Suspension in Y-arm\\
\midrule
EXPUM& L1:SUS-ETMX\_L2\_WIT\_L\_DQ&			BSC4&	End X-arm Penultimate Mass\\
\bottomrule
\end{tabular}
\label{tab:scat_list}
}
\end{center}
\end{table}

\pagebreak

\section{Seismic Isolation and Suspensions Layout of Advanced LIGO} 
\label{ap:layout}
\begin{figure}[!ht]
	\begin{center}
		\includegraphics[trim={0 0 10.5cm 0}, width=0.8\columnwidth]{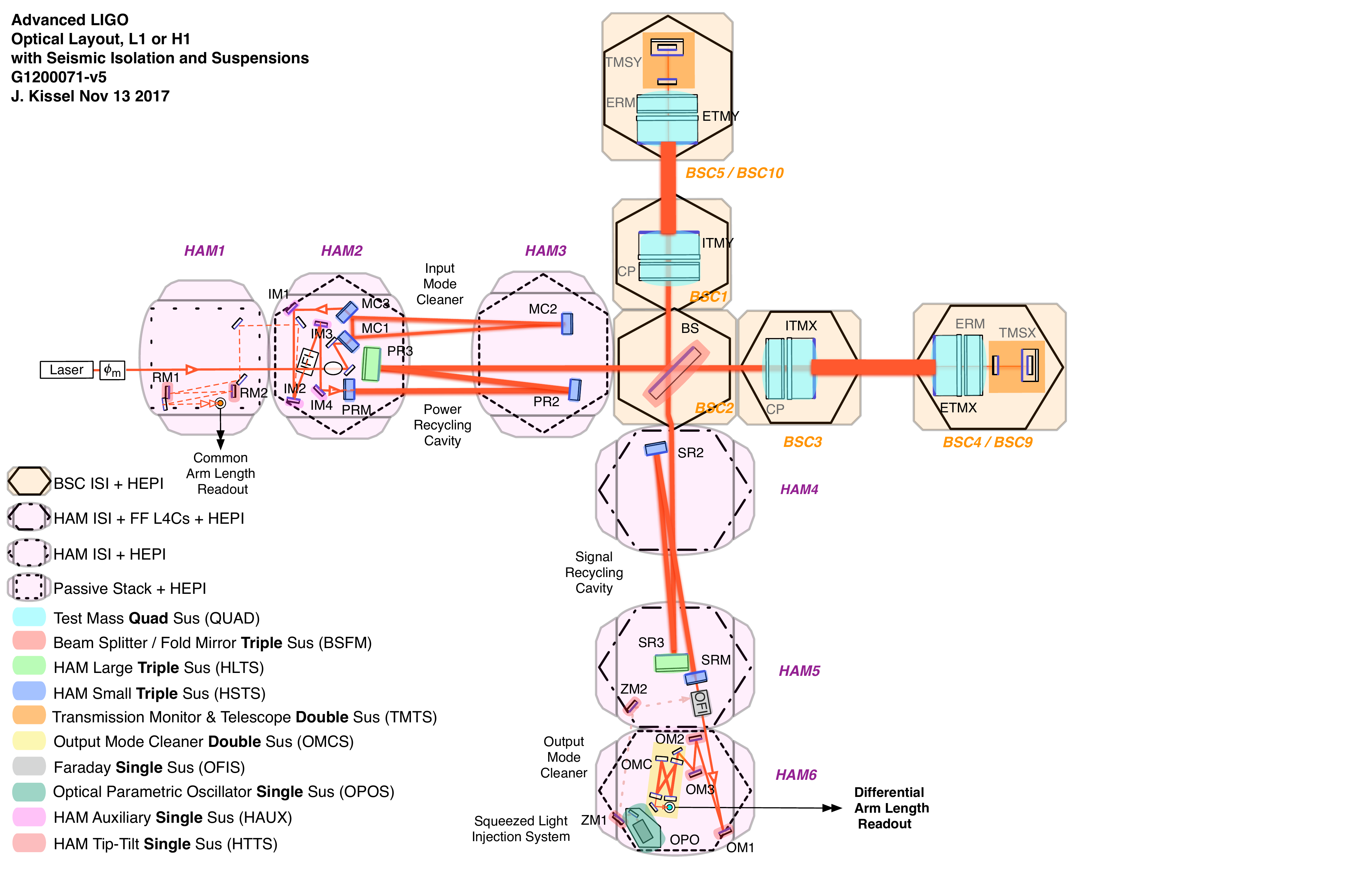}
		\caption{
		Schematic of the Advanced LIGO optical layout \cite{Kissel_2017} showing the laser beam path through optics, which are suspended via many different types and different number of layers of seismic isolation. BSC chambers house the core interferometer optics: the beam splitter, two sets of ``input'' and ``end'' test masses that comprise each Fabry-Perot arm cavity. These test mass quadruple pendula (QUAD) contain two parallel suspension chains, one for the test mass itself, and one for a ``reaction chain’’ where each stage is used for quiet actuation of the corresponding stage of the main chain. The transmission monitor and telescope suspension (TMTS) holds optics used for arm cavity power monitoring and angular displacement sensing during observation and arm length stabilization during lock acquisition. HAM chambers house auxiliary optics and platforms used for power recycling, signal recycling, beam expanding or reducing telescopes, mode cleaning, beam steering, supporting the squeezed light system, and suspending the differential arm length readout used for gravitational wave detection. For a review of these systems, see for example \cite{matichard2016overview}, and references therein.}
	\end{center}
	\label{fig:cartoon}
\end{figure}


\end{document}